\documentclass{iopart}

\usepackage{epsfig}
\newcommand{\beq}{\begin{equation}}
\newcommand{\eeq}{\end{equation}}

\newcommand{\as}{\alpha_s} 
\newcommand{\un}{\underline}

\begin{document}
\title{Heavy quark production from Color Glass Condensate at RHIC}
\author{K.L. Tuchin}
\address{ Physics Department, Brookhaven National Laboratory,\\
Upton, NY 11973-5000, USA}
\ead{tuchin@quark.phy.bnl.gov}
\begin{flushright}
{\small BNL--NT--04/6}
\end{flushright}

\begin{abstract}
I consider production of heavy quarks in pA and AA
collisions in the framework of the Color Glass Condensate. I discuss  
the heavy quark production in pA collisions in a classical approximation 
which takes into account all multiple rescatterings of a proton in a 
nucleus. For peripheral collisions, heavy quark production cross section 
can be written in $k_T$-factorized form. The $k_T$ factorization is used 
to construct a simple model which takes into account both classical and  
quantum effects in pA and AA collisions. I review the main result of 
calculation based on that model: open charm in the central rapidity region 
at RHIC gets suppressed as a function of rapidity. Although the 
numbers obtained for the suppression factor are model dependent, the very 
fact of suppression is the general  feature of the Color Glass Condensate 
at RHIC kinematical region. It indicates the onset of quantum evolution 
effects in a nucleus and is universal for all inclusive processes.

\end{abstract}

The process of  heavy quark production in pA collisions at high energies  
has three separated in time stages in the
nucleus rest frame. Emission of a gluon $g$ by a proton's
valence quark $q_v$ takes much longer time $\tau_{q_v\rightarrow q_vg}$
than a
subsequent emission of a $q\bar q$ pair by a gluon $ t_{g\rightarrow
q\bar q} \ll \tau_{q_v\rightarrow q_vg}$. In turn, the time of
interaction of a $q_vgq\bar q$ system  with
a nucleus is of the order of nuclear length $R_A$ and is negligible as
compared to the evolution time of the proton wave function
$\tau_{q_v\rightarrow q_vg}\gg t_{g\rightarrow  q\bar q}\gg R_A$.
For example, the gluon emission takes time 
\beq
\tau_{q_v\rightarrow qg}\approx \frac{2 q^+}{\un q^2}=\frac{1}{x M_N}\gg 
R_A\simeq \tau_\mathrm{int},
\eeq
where the proton is moving in $+$ light-cone direction. 
The same argument applies to the successive emission of a quark-antiquark
pair by a gluon in a proton's wave function. The separation of 
times is ensured by strong ordering $q^+\gg k^+\gg k_T$, where 
$k^+$ and $k_T$ are the quark energy and transverse momentum respectively. 
Therefore, processes in which gluon or heavy quarks are produced in 
course of the 
rescatterings in a nucleus are suppressed by powers of energy $p^+$ 
\cite{ioffe}.

As the result, the differential cross section for the quark production can 
be written in a factorized form as a convolution of the valence quark 
and gluon wave functions with the rescattering factor. The final results 
takes form \cite{KopTar,Tu,BGV}
\begin{eqnarray}\label{main}
\fl\frac{d\sigma}{d^2k \,dy}=
\int d^2b\, \int d^2x_0
\int d\alpha\,
\int \frac{d^2x\, d^2 y}{(2\pi)^3}\, \Phi_\mathrm{g\rightarrow
q\bar q}(\un x,\un x_0,\un y, \alpha)\,
\Phi_\mathrm{q_v\rightarrow
q_v\bar g}(\un x,\un x_0,\un y, \alpha)\,   
\nonumber\\
\fl\times e^{-i\un k\cdot (\un x-\un y)}\,
\bigg(1+e^{-\frac{1}{4}\, \frac{C_F}{N_c}\, (\un x-\un y)^2\, Q_s^2}-
e^{-\frac{1}{4}\, \frac{C_F}{N_c}\, (\un x-\un x_0) ^2\, Q_s^2}-
e^{-\frac{1}{4}\, \frac{C_F}{N_c}\, (\un y-\un x_0)^2\, Q_s^2}\bigg)\,,
\end{eqnarray}
where I assumed for simplicity that the dominant contribution comes from 
interaction of the $q\bar q$ pair with the target, while rescatterings of 
the gluon and the valence quark are neglected. However in general, they 
must be taken into account as well. In Eq.~(\ref{main}) I used the 
following 
notations: 
\beq\label{wf1}
\fl\Phi_\mathrm{q_v\rightarrow
q_v\bar g}(\un x,\un x_0,\un y, \alpha)=
\frac{\as\, C_F}{\pi^2}\,
\frac{(\alpha\un x+(1-\alpha)\un x_0) \cdot(\alpha\un y+(1-\alpha)\un x_0)}
{(\alpha\un x+(1-\alpha)\un x_0)^2\,(\alpha\un y+(1-\alpha)\un x_0)^2}\,,
\eeq
\begin{eqnarray}\label{wf2}
\fl\Phi_{g\rightarrow q\bar q}(\un z,\un x, \un x_0, \alpha)=
\frac{\as}{\pi}\, m^2\,\bigg(\, \frac{(\un x-\un x_0)\cdot
(\un y -\un x_0)}{|\un x-\un x_0|\, |\un y -\un x_0|}   
K_1(|\un x-\un x_0|\,m)\,K_1(|\un y -\un x_0|\,m)\nonumber\\
\fl\times[\,\alpha^2\,+\,(1-\alpha)^2\,]
+\,K_0(|\un x-\un x_0|\,m)K_0(|\un y -\un x_0|\,m) \,\bigg)\,,
\end{eqnarray}
where $Q_s$ is the saturation scale, $m$ is a quark mass and
$\alpha=k^+/q^+$.

In the nucleus light-cone frame Eq.~(\ref{main}) can be interpreted as a 
cross 
section for proton scattering off the $q\bar q$ fluctuation of the strong 
nuclear gluonic field (Color Glass Condensate). At the classical level, 
this field is of the order $F\sim Q_s^2/g$. It is well known in QED that 
strong electric field $E$ can produce quark-antiquark pairs from vacuum if 
change of its energy $eE$ along distance $\hbar /mc$ is larger than 
$mc^2$ i.\ e.\ $eE>m^2$. Similarly, strong chromoelectric field can produce 
pairs from vacuum if $Q_s^2>m^2$ \cite{KhT}. Theory of Color Glass 
Condensate 
predicts dependence of the saturation scale on energy and atomic number: 
$Q_s\propto A^{1/6} s^{\alpha_s}$ \cite{LT}. Thus, one expects that the 
production 
pattern of heavy quarks is the same as light ones, once the 
collision energy is high enough.

\begin{figure}
\begin{center} 
\begin{tabular}{ll}
\epsfig{file=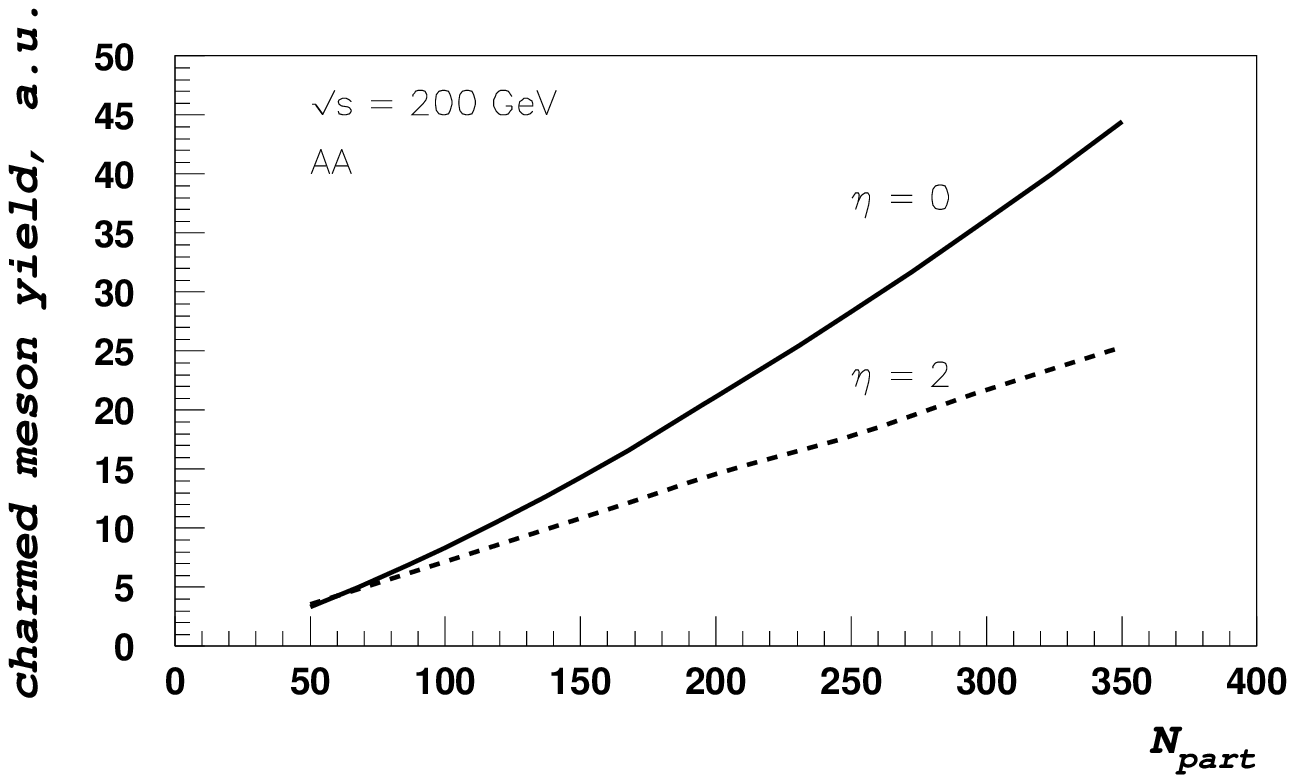, width=7cm}&
\epsfig{file=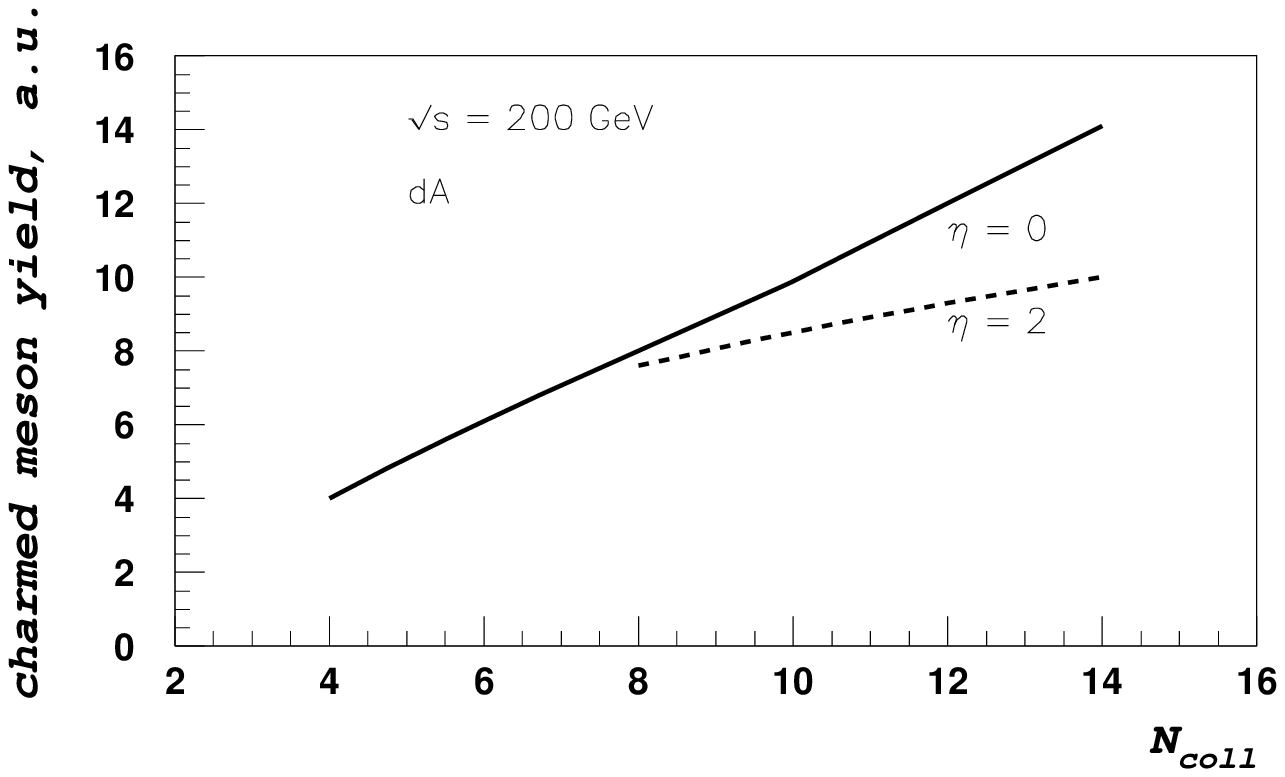,width=7cm}
\end{tabular}  
\caption{Dependence of the charmed meson yield on centrality at 
central  and forward rapidities for dAu and AuAu 
collisions. 
}
\label{fig1}
\end{center}
\end{figure}

\begin{figure}
\begin{center}
\epsfig{file=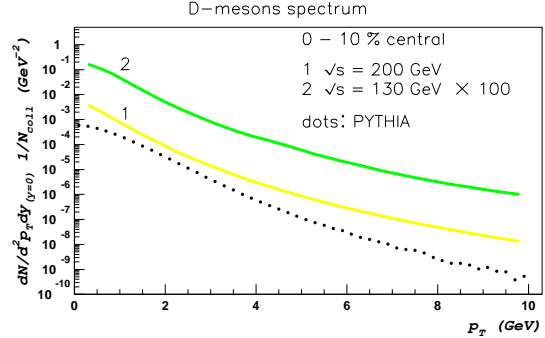, width=8cm}
\caption{ $\frac{1}{2}(D^0+\bar D^0)$ spectra in heavy-ion collisions at  
$\sqrt{s}=130$ GeV and $\sqrt{s}=200$ GeV. 
}\label{fig2}
\end{center}
\end{figure}

Unfortunately, there is no calculation of the heavy quark 
production cross section for the AA reaction. Therefore, to get a 
numerical results which can be compared with RHIC experiments one uses 
phenomenological models which take into account the important features of 
the heavy quark production by strong color fields. This can be done by 
expanding exponents in Eq.~(\ref{main})(together with terms describing 
gluon rescattering omitted here) to the leading order in $Q_s^2$  
and Fourier-transforming to the momentum space. This way one gets the 
cross section for the quark production in pp collisions in 
$k_T$-factorized form \cite{LRSS} 
\beq\label{kT}
\fl\frac{d\sigma}{d^2p\,dy_1\,dy_2}=
\int d^2q_1\int d^2q_2\,\phi(x_1,\un q_1)\,\mathcal{A}(\hat s,\hat t, \hat 
u, \un q_1^2, \un q_2^2)\,\phi(x_2,\un q_2),
\eeq
where $\mathcal{A}$ is the amplitude for the subprocess process $g^*(\un 
q_1)g^*(\un q_2)\rightarrow ggq\bar q$. It reduces to the familiar 
perturbative amplitude for on-shell gluons $gg\rightarrow q\bar q$ after 
taking the limit $\un q_1^2,\un q_2^2\rightarrow 0$ and
averaging over azimuthal angles of $\un q_1$ and $\un q_2$. However, this 
approximation turns out to be not valid at high energies since it 
disagrees with the experimental data on charm and bottom production at 
Fermilab.
$\phi(x,\un q)$ in (\ref{kT}) is the unintegrated gluon distributions such 
that $\phi(x,\un q)=d\, xG(x,\un q)/d \un q^2$. 
In principle $\phi(x,\un q)$ can be found 
from the QCD evolution equations at high energy \cite{BK}. However, 
finding the numerical solution to those equations is still not completely 
solved problem even in the mean-field approximation. We rather prefer to 
use a simple model for the unintegrated gluon distributions based on 
available theoretical information about the solution to the evolution 
equations. 
Assume that the cross section for the quark production in pA and AA 
reactions can be written in the same $k_T$ factorized form as for pp case 
(\ref{kT}). Then,  $\phi(x,\un q)$ must satisfy the following 
conditions:
\beq\label{sat}
\phi(x,\un q^2)=\frac{4\,C_F\,\pi\, R_A^2}{\as\, 2\, (2\pi)^2}\,
\ln (Q_s^2/\un q^2)\,,\quad \un q^2<Q_s^2,\,\, \mathrm{saturation}
\eeq
\beq\label{pert}
\phi(x,\un q^2)\sim \un q^{-2},\quad  q^2> k_\mathrm{geom}^2,\,\,
\mathrm{pQCD\,\, behavior}
\eeq
where $k_\mathrm{geom}^2\gg Q_s^2$ is the ``extended geometric scaling"
\cite{IIM} momentum squared, and 
\beq\label{ext}
\phi(x,\un q^2) = \phi(\un q^2/Q_s^2)\,,\quad \un 
q^2<k_\mathrm{geom}^2,\,\,\mathrm{``extended\,\, geometric\,\, scaling"}
\eeq
Of course there is more than one function satisfying these conditions.
In my paper with D.~Kharzeev \cite{KhT} a parametrization is 
suggested which is inspired by the double logarithmic approximation to the 
DGLAP and BFKL equations. Our numerical calculations have been done with 
that function. The results of calculations are presented in 
Fig.~\ref{fig1} and Fig.~\ref{fig2}.

The charmed meson yield shown in Fig.~\ref{fig1} gets suppressed as one 
goes from central rapidity $y=0$ to forward rapidity $\eta=2$ both in pA 
and AA collisions. This is because the saturation scale of a nucleus 
increases 
with rapidity. At RHIC $Q_s^2\simeq 2\,e^{0.3 y}$~GeV$^2$. Therefore,
at midrapidity $Q_s^2<m^2_c$ and the charmed quarks are 
produced perturbatively. As the result yield scales like 
\beq\label{mid}
\frac{dN_{AA}}{d\eta}\big|_{\eta=0}\sim A^{4/3}\sim N_\mathrm{coll}\,.
\eeq
At forward rapidity $Q_s^2>m^2_c$ implying that
\beq\label{for}
\frac{dN_{AA}}{d\eta}\big|_{\eta\ge 2}\sim A\sim N_\mathrm{part}\,.
\eeq
The same formulas hold for dA after replacement of $N_\mathrm{part}$ by 
$N_\mathrm{part}^{Au}$ and $N_\mathrm{coll}$ by
$N_\mathrm{coll}^{Au}$.

The suppression effect is insensitive to high 
energy quantum evolution effects  which change the anomalous dimension of 
the gluon distribution functions for transverse momenta satisfying 
$Q_s^2<\un q^2<k_\mathrm{geom}^2$. If however one applies an infrared   
cutoff on a spectrum at transverse momenta of the order of $Q_s$, then 
the region $\un q^2<k_\mathrm{geom}^2$ where the quantum effects are 
important will dominate the yield. The resulting nuclear modification 
factors scale like $R_{AA}\sim 1/A^{1/3}\sim 
1/N_\mathrm{part}^{1/3}\simeq 0.5$ and
$R_{dA}\sim 1/A^{1/6}\sim
1/\sqrt{N_\mathrm{part}^{Au}}\simeq 0.75$ in the forward region. As we 
have already noted, these suppression factors at forward rapidities are of 
the same order of magnitude as in inclusive meson production case, since 
at low-$x$ (forward rapidities) production pattern of charmed quarks is 
the same as the light ones. 

In  Fig.~\ref{fig2} we compare the open charm spectrum obtained from our 
saturation model based on $k_T$-factorization with predictions based on 
collinear factorization approach (PYTHIA).  
We observe that PYTHIA spectrum is significantly softer than ours.
Therefore, we predict much harder open charm spectrum. Of
course, the steepness of the PYTHIA spectrum depends on the intrinsic 
momentum parameter $k_0$. Our spectrum can probably be reproduced by 
PYTHIA if one takes $k_0\simeq Q_s\gg \Lambda_\mathrm{QCD}$ and fixes the
$K$-factor. This large value of intrinsic $k_T$ would however signal
the breakdown of the collinear factorization approach.

\ack{The author is indebted to Dmitri Kharzeev for collaborating in most 
of the results discussed in this publication.
This research was supported by the U.S. Department of
Energy under Grant No. DE-AC02-98CH10886.
}

\end{document}